\documentclass[12pt]{article}
\usepackage{epsf}
\usepackage{graphicx}
\usepackage{amssymb}
\usepackage{amsmath}
\usepackage{color}
\usepackage[utf8]{inputenc}
\usepackage[english]{babel}

\begin{document}
\title*\begin{center}
{\textbf{{\LARGE The Size Seems to Matter or Where Lies the
"Asymptopia"?}}}
\end{center}

\begin{center}
V. A. Petrov
\end{center}
\begin{center}
Division of Theoretical Physics, A. A. Logunov Institute for High Energy
Physics,
\begin{center}
NRC "Kurchatov Institute", Protvino, RF
\end{center}
\end{center}

\begin{center}
V. A. Okorokov
\end{center}
\begin{center}
Department of Physics, National Research Nuclear University MEPhI
\end{center}
\begin{center}
(Moscow Engineering Physics Institute), Moscow, RF
\end{center}

\begin{center}
Abstract
\end{center}
\textit{We discuss an apparent correlation between the onset of
the rising regime for the total cross-sections and the slowdown of
the rise of the forward slopes with energy. It is shown that even
at highest energies achieved  with the LHC the proper sizes of the
colliding protons comprise the bulk of the the interaction
region. This seems to witness that the "asymptopia" -- a
hypothetical "truly asymptotic" regime -- lies at energies no less
than $\mathcal{O}$(100 TeV). In the course of reasoning we also discuss the question of the dependence of the effective sizes of hadrons in collision on the type of their interaction.}

\section*{The Problem} \label{sec:0}
\begin{figure}[!h]
\begin{center}

\includegraphics[width=12.0cm,height=13.0cm]{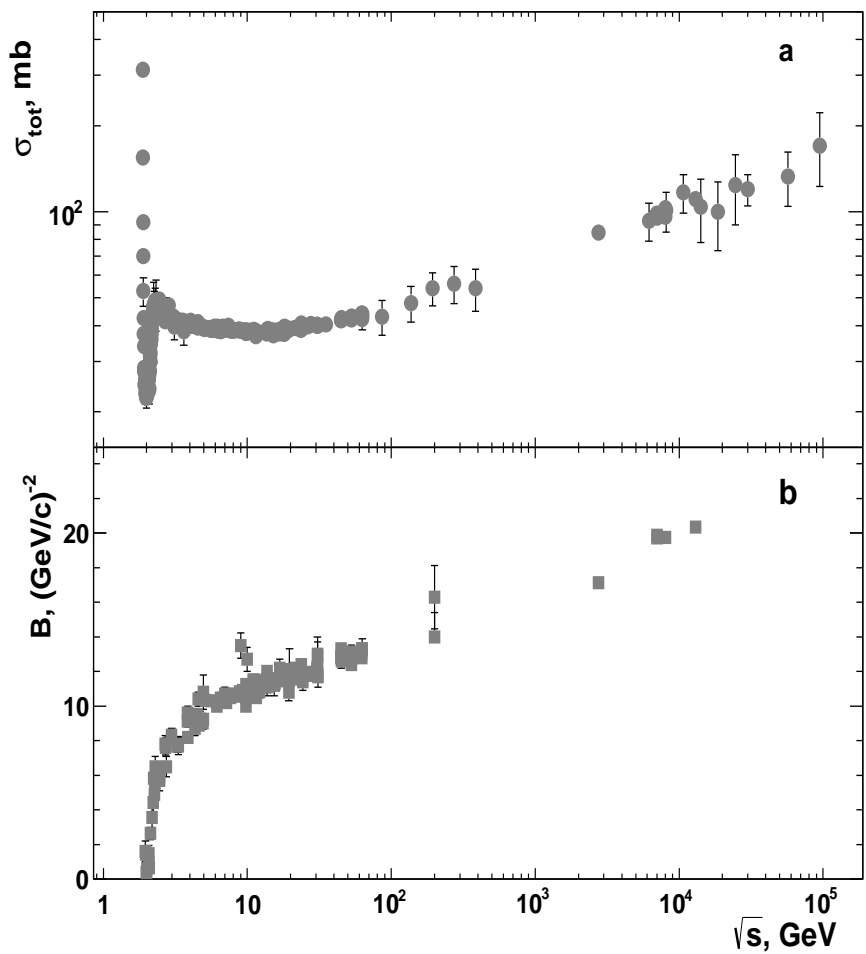}
\end{center}
\caption{\textit{Total cross-sections (a) and forward slopes (b)
as functions of the $pp$ collision energy. Experimental results are from DB17+.}} \label{fig:1-2017}
\end{figure}

Let us look at Fig. \ref{fig:1-2017} presenting the energy
evolution of the total cross sections
($\sigma_{\footnotesize\mbox{tot}}$) and forward slopes ($B$) for
proton-proton interactions. The database (DB17+) of
experimental results for the set of the scattering parameters
$\mathcal{G}_{pp} \equiv
\{\mathcal{G}_{pp}^{i}\}_{i=1}^{2}=\{\sigma_{\footnotesize\mbox{tot}},B\}$
is used in the present paper\footnote{In the paper total errors
are used for experimental points unless otherwise specified. The
total error is calculated as systematic error added in quadrature
to statistical one.}. The set for
$\sigma_{\footnotesize\mbox{tot}}$ contains the data from
\cite{Okorokov-IJMPA-32-1750175-2017} and preliminary TOTEM results
at $\sqrt{s}=2.76$ \cite{Nemes-PoS-DIS2017-059} and 13 TeV
\cite{Antchev-arXiv-1712.06153}, the data sample for $B$ unites the
subset from \cite{Okorokov-AHEP-2015-914170-2015} with recent
improvement from \cite{Antchev-EPJC-76-661-2016} and the new result
at $\sqrt{s}=8$ TeV \cite{Aabound-PLB-761-158-2016} as well as
with preliminary points at $\sqrt{s}=0.20$
\cite{Koralt-PhD-2013}, 2.76 \cite{Turini-Report-LHCC-22022017}
and 13 TeV \cite{Antchev-arXiv-1712.06153}.

It is seen by eye that the total cross-section starts to rise
approximately in the same energy interval ($\sim$ 10--11 GeV) where the ever rising
forward slope slows down. What could it mean? Is there some
fundamental mechanism underlying both phenomena or this is just a
hazardous coincidence?

$S$-matrix acts on the Hilbert space spanned by the asymptotic
Fock states of free particles and to describe all possible
processes occurring after collision of two initial particles the
$S$-matrix elements as functions of relevant momenta are fairly
sufficient. Nonetheless, the question: "What happens between the
two, \textit{in}- and \textit{out}-, spatio-temporal infinities?"
arises again and again. Since the Yukawa discovery it was clear
that there exists a region of space where one cannot discern among
separate in- or outgoing particles. Its extent should give the
information about the forces causing the scattering. The problem
is how can we, dealing only with the data from the detectors
asymptotically remote from this region both in space and time,
know something about it? In classical mechanics the scattering
cross-sections relate the scattering angle with the impact
parameter. In quantum mechanics such a direct relation is lost:
due to the uncertainty relations all impact parameters contribute.
However in the sense of average quantities we can still extract
some information about "effective" impact parameters and, hence,
the spatial extent of the interaction potential. At relativistic
velocities of the colliding particles the instantaneous potential
ceases to be adequate. Retardation effects can spoil the image of
the scatterer which is no longer a snapshot but rather some kind
of average in time.

With all this in mind, we are going to consider now the main observable
characteristics of the elastic scattering of hadrons related to
spatial scales which are mostly in use in experimental studies and
try to summarize both general and model description and properties
of them.
We will see that spatial measures can more or less unambiguously witness if we are in the "truly asymptotic region", "Asymptopia", when in the LHC energy region.

 It seems natural, if to discuss spatial scales, to start from the "proper sizes" of colliding particles. In literature one can find a number of papers devoted to or seriously concerned with spatial characteristics ("sizes") of hadrons in collision.
Not being able to properly comment on majority of them here, we still would like to give them a tribute of our respect \cite{huf}.

 In what follow we  shall discuss nucleons but the essential part of our arguments fairly concerns generic hadrons.

\section*{1. Nucleon Size} \label{sec:1}
Modern view of the nucleon is, roughly, a valence quark core
immersed into the parton "sea" of virtual $q\bar{q}$ pairs and
gluons. Such a picture depends on the Lorentz frame as different
quantum fluctuations live different times according to the
energy-time uncertainty relations. At rest an unpolarized nucleon
can be considered as some fuzzy ball the average size of which is
defined by the valence quarks. If to consider the nucleon in the
frames where it moves faster and faster then the role of the
vacuum fluctuations gets dominant and its transverse size becomes
an asymptotically universal function of its energy. At which
energy this dominance overwhelms is an interesting question but has no a definite answer at the moment.

How is it
possible to measure the nucleon size? The answer can be done if we mean the valence quark core of the nucleon.
The standard way to estimate the so understood nucleon size is to measure its
electromagnetic form factor, $F(t)$, to find the best fitting of it as a function of
the transferred momenta and to extract the "charge radius"
$r_{ch,N}$ according to the well-known formula
\begin{equation}
\displaystyle r^{2}_{ch,N} =
6\left.\frac{dF(t)}{dt}\right|_{t=0}. \label{eq:1.1}
\end{equation}

In which way this formula gives us the nucleon size independent on peculiarities of its probe (electromagnetic in this case)?  The best would be to
measure the gravitational form-factors of nucleons as they give us
(in a universal way)the matter distribution inside the nucleons.
Unfortunately, this is still inaccessible. So we are enforced to take the standard way using the quantities like (\ref{eq:1.1}).
However, the direct interpretation of this formula for the neutron leads to absurd: $r^{2}_{ch,n}$ is \textit{negative}. This is
implied by the fact that electromagnetic form factors are related
to the charge densities which can have any sign and are not
directly related to the matter densities (which are positive
defined).

Nonetheless, we still are able to use "charge radii" to extract the physical nucleon size.
We take use of the fact that the carriers of electric charge of the nucleon, i.e.
valence quarks $u$, $d$ are at the same time the sources of strong
interaction forces as they are coupled to gluon fields, the basic
agent of strong interactions. We will assume that the "sea" of $q\bar{q} $ pairs and gluons is the result of the QCD  vacuum
polarization by valence quarks. Exactly this process of "parton
(allegedly, gluon) diffusion" \cite{Gr} forms the  main agent of the strong interaction,
the celebrated Pomeron, which stipulates the leading contribution in high-energy dependence of
the interaction region, quantified by $B(s)$, and
cross-sections. Being related to vacuum, the Pomeron mechanism is
universal for all hadrons, independently of their valence
structure. Additional forces between the nucleons are generated by
the valence quark interchanges and (in case of $p\bar{p}$
collisions) their annihilation. Contrary to the Pomeron, these
forces, quantified by "secondary Reggeons", strongly depend on the
valence content of colliding hadrons but die-off at high enough
energies (i.e. starting from the ISR). So we believe that the
average proper size of the nucleon , i.e. the size of the region
where the sources of strong interaction are concentrated, is given
by those of the valence quarks, $u$ and $d$.

Now we are to extract the nucleon proper size from the proton and neutron form factors.
In the spirit of the above said the proton and neutron electric form factors are related to the valence quark distributions in the following way
\begin{center}
$\displaystyle F_{p}(t) = \frac{2}{3}\int dx u_{p}(x,t) -
\frac{1}{3}\int dx d_{p} (x,t)$,
\end{center}
\begin{center}
$\displaystyle F_{n}(t) = -\frac{1}{3}\int dx d_{n}(x,t) +
\frac{2}{3}\int dx u_{n}(x,t)$.
\end{center}
Here
\begin{center}
$\displaystyle u_{p}(x,t)= \int
d^{2}bJ_{0}(b\sqrt{-t})\tilde{u}_{p}(x,\textbf{b}) $
\end{center}
where  $t= -\textbf{q}^2$ while $\textbf{q}$ is the 2D vector conjugated to $\textbf{b}$. This introduces the quantity  $\tilde{u}_{p}(x,\textbf{b}) $ which  is the valence $u$-quark
number density in the proton in the longitudinal momentum fraction
$ x $ and the transverse position of the quark $\textbf{b}$
relatively to the center of the proton \cite{Ki}. Normally this center is defined as
the origin in the frame where
$$ \sum_{j\in p} x_{j}\textbf{b}_{j} = 0.$$
Similarly for other parton functions. At $ t=0 $ we get usual valence quark densities
measured in DIS:
\begin{center}
$u_{p}(x,0)=u_{p}(x),~~~~~d_{p}(x,0)=d_{p}(x)$
\end{center}
and similarly for the neutron.
Evidently
\begin{center}
$\displaystyle \int d^{2}b\,dx \tilde{u}_{p}(x,\textbf{b}) = \int
dx {u}_{p}(x) = 2$ ,
\end{center}
\begin{center}
$\displaystyle \int d^{2}b\,dx \tilde{d}_{p}(x,\textbf{b}) = \int
dx {d}_{p}(x) = 1 .$
\end{center}
Similarly for the neutron. Here we do not explicitly indicate the
renormalization scale dependence of quark densities. We only note
that RG non-singlet evolution of valence quark densities in no way
contradicts the RG-invariance of the nucleon form factors.
Isotopic invariance implies that
\begin{center}
$d_{n} = u_{p},~~~~~u_{n} = d_{p}$.
\end{center}
From this relations we obtain
\begin{center}
$\displaystyle F_{p}(t) = \frac{2}{3}\int dx u_{p}(x,t) -
\frac{1}{3}\int dx d_{p}(x,t)  $,
\end{center}
\begin{center}
$\displaystyle F_{n}(t) = -\frac{1}{3}\int dx u_{p}(x,t) +
\frac{2}{3}\int dx d_{p}(x,t)$.
\end{center}
This relation give us an opportunity to extract the nucleon proper
size from the data on the proton and neutron form factors.

Indeed, the average positions of $u$- and $d$- quarks are given by
the formula
\begin{center}
$\displaystyle 2\bigl\langle r^{2}_{u}\bigr\rangle =
6\biggl[\frac{d}{dt}\displaystyle \int dx
u_{p}(x,t)\biggr]_{t=0}$,
\end{center}
\begin{center}
$\displaystyle \bigl\langle r^{2}_{d}\bigr\rangle =
6\biggl[\frac{d}{dt}\displaystyle \int dx
d_{p}(x,t)\biggr]_{t=0}$.
\end{center}
Thereof we easily come to the expression of the valence quark
average positions in terms of observed "charge radii" of the
proton and neutron:
\begin{center}
$\displaystyle \bigl\langle r^{2}_{u}\bigr\rangle = r^{2}_{ch,p}+
\frac{1}{2}r^{2}_{ch,n}$,
\end{center}
\begin{center}
$\bigl\langle r^{2}_{d}\bigr\rangle = r^{2}_{ch,p} + 2
r^{2}_{ch,n}$.
\end{center}
PDG \cite{pdg-2016} gives
\begin{center}
$r^{2}_{ch,n} = - 0.1161 \pm
0.0022~\mbox{fm}^{2}$
\end{center}
while for the proton there are two values, "the proton radius puzzle":
\begin{center}
$ r_{ch,p}\,(\mu p~\mbox{Lamb shift})=0.8409 \pm
0.0004~\mbox{fm}$
\end{center}
and
\begin{center}
$r_{ch,p}\,(ep~\mbox{CODATA  value})= 0.875 \pm
0.006~\mbox{fm} $.
\end{center}
Let us first extract the quark positions in the proton.

The $ \mu p~\mbox{Lamb shift} $ option gives:
\begin{center}
$\displaystyle \bigl\langle r^{2}_{u}\bigr\rangle = (0.8056 \pm 0.0011 ~\mbox{fm})^{2}$,~~~
$\bigl\langle r^{2}_{d}\bigr\rangle = (0.6891 \pm 0.0017 ~\mbox{fm})^{2} $
\end{center}
while the CODATA leads to
\begin{center}
$\displaystyle \bigl\langle r^{2}_{u}\bigr\rangle = (0.872 \pm 0.006~\mbox{fm})^{2}$,~~~
$\bigl\langle r^{2}_{d}\bigr\rangle = (0.731 \pm 0.008~\mbox{fm})^{2} $.
\end{center}
Note that both options give practically the same excess ($\sim$1.2) of the $u$-quark position over the $d$-quark one.

This qualitatively corresponds to a slightly heavier $d$-quark. Just for fun, if we take the arithmetic average of the mass ratios
$\bigl\langle m_{d}/m_{u}\bigr\rangle $ which, according to PDG \cite{pdg-2016} is $\approx 2.18$, then with an acceptable accuracy
\begin{center}
$\bigl\{\bigl\langle r^{2}_{u}\bigr\rangle / \bigl\langle r^{2}_{d}\bigr\rangle\bigr\}~\mbox{(CODATA)}
= \bigl[\bigl\langle m_{d}/m_{u}\bigr\rangle \bigr]^{1/2}$.
\end{center}
Note that the PDG values of the (running) quark masses are referred to the scale of order 2 GeV. Nonetheless, due to the fact that QCD evolution does not mix different flavours the ratio $ m_{d}/m_{u} $ is scale independent.
We, however, are not going to develop further this observation in the present article.

 Now, it is turn of the nucleon size (the same for the proton and the neutron in the approximation of exact isotopic symmetry) defined as
\begin{center}
$\displaystyle \bigl\langle r^{2}_{N}\bigr\rangle = \frac{2}{3} \bigl\langle r^{2}_{u}\bigr\rangle + \frac{1}{3} \bigl\langle r^{2}_{d}\bigr\rangle = r^{2}_{ch,p}+r^{2}_{ch,n}$.
\end{center}
Note that the coefficients 1/3 and 2/3 before the quark average sizes relate to corresponding probabilities and not to the quark charges. Again, the $ \mu p~\mbox{Lamb shift} $ option gives:
\begin{center}
$\displaystyle \bigl\langle r^{2}_{N}\bigr\rangle = (0.7687\pm 0.0015~\mbox{fm})^{2}$
\end{center}
and the CODATA value is
\begin{center}
$\displaystyle \bigl\langle r^{2}_{N}\bigr\rangle = (0.806 \pm 0.012~\mbox{fm})^{2}$.
\end{center}
Keeping these values in mind, we shall estimate the proton size from the
forward slope data and compare it with the above competing values.

\section*{2. Basic scattering observables } \label{sec:2}
We now come proper to the proton-proton scattering and need to fix terms and designations.
Fundamental element of everything what follows is the scattering
amplitude
\begin{center}
$T(s,t)=|T(s,t)|\exp i \Phi(s,t)$.
\end{center}
Being the observable quantity, the differential cross section
\begin{center}
$\displaystyle \frac{d\sigma}{dt}= \frac{1}{16\pi
s(s-4m^2)}|T(s,t)|^2$
\end{center}
depends on its modulus only. It doesn't mean that the scattering
phase $\Phi(s,t)$ absolutely defies measurements. Fortunately,
besides the strong interaction there is the electromagnetic one.
Suppose for the sake of simplicity that the full scattering
amplitude is just the sum of the strong interaction amplitude, $
T_{s}(s,t) $, and that of the electromagnetic interaction,
"Coulomb", $T_{C}(s,t)$. In the lowest order in the fine structure
constant the latter has no phase and can be considered as a known.
Differential cross-section contains now the interference term
\begin{center}
$2|T_{s}(s,t)| T_{C}(s,t)\cos\Phi(s,t)$
\end{center}
and one could try to extract the strong phase from the data
but\dots one needs to know $|T_{s}|$! So the modulus and the
phase of the strong interaction amplitude cannot be separately
measured in a model-independent way. This sad fact in no way
confuses physicists and the quantity
\begin{center}
$\displaystyle \rho(s)\equiv\frac{{\rm Re\,T}(s,0)}{{\rm
Im\,T}(s,0)} = \cot \Phi(s,0)$
\end{center}
is considered as one of the \textit{bona fide} basic observables. Actually, the problem is much more complicated and we refer the interested reader to \cite{CNI} where the issues of Coulomb-nuclear interference are reviewed and discussed.

Just for completeness we mention the total cross-section
\begin{center}
$\displaystyle \sigma_{\footnotesize\mbox{tot}}  = \frac{1}{2i\sqrt{s(s-4m^{2})}} \lim_{t\rightarrow 0} \bigl[T(s+i\varepsilon, t) - T(s-i\varepsilon, t)\bigr]$
\end{center}
and elastic cross-section
\begin{center}
$\displaystyle \sigma_{\footnotesize\mbox{el}} = \int dt \frac{d\sigma}{dt}.$
\end{center}
At last, two more characteristics which are  being discussed in this paper,  are the \textit{local (logarithmic) slope}
\begin{center}
$\displaystyle B(s,t) \equiv \frac{1}{d\sigma / dt}\frac{\partial
[d\sigma / dt]}{\partial t} = \frac{\partial \ln[d\sigma /
dt]}{\partial t}$
\end{center}
and the \textit{diffraction peak width} (now a bit out-of-mode)
\begin{center}
$\displaystyle \tilde{\Delta}^{-1} \equiv
\frac{1}{\sigma_{\footnotesize\mbox{el}}}\biggl[{\frac{d\sigma}{dt}}\biggr]_{t=0}
.$
\end{center}

\section*{3. Physical meaning } \label{sec:3}

Let's now try to understand which physical meaning, besides their
formal definition, bear these characteristics. Let's start from the
phase. In quantum mechanics one has the following relation
\begin{center}
$\displaystyle \langle x_{i}\rangle =\biggl\langle
\frac{\partial\varphi (p)}{\partial p_{i}}\biggr\rangle$,
\end{center}
where $\varphi (p)$ is the phase of the wave function in the momentum
space.

The scattering amplitude, duly normalized, gives the
probability amplitude for the momenta of the scattered particles.
This allows us to operationally define the average value of the difference of
the $i$-th coordinate component of the scattered particles. When using the term "coordinate" we always mean the coordinate of its center of mass as we deal with extended particles. E.g., we
obtain for the longitudinal coordinate (Newton--Wigner type modification
of the position operator quickly dies off with energy)  distance
between the outgoing particles in the c.m.s. frame
\begin{center}
$\displaystyle \langle x_{\parallel}\rangle =\biggl\langle
\frac{\partial\Phi}{\partial p_{\,\parallel}} \biggr\rangle =
\sqrt{s-4m^2}\biggl\langle \frac{\partial\Phi(s,t)}{\partial
t}\biggr\rangle $
\end{center}
and the average in the last term is taken with
$\sigma_{\footnotesize\mbox{el}}^{-1}d\sigma / dt$. We have to
emphasize again that it comes to the coordinate of scattered
particles when they, loosely speaking, leave the interaction region
and not when they reach remote detectors. We see that the
knowledge of the scattering phase could give a very important
information about the spatial extent of the interaction region.

The local slope, by definition, signals about the change of the
slope of the $t$-distribution. For instance, one can write
\begin{center}
$\displaystyle \frac{d\sigma}{dt}=\frac{d\sigma}{dt}(s,0)\exp
[t\hat{B}(s,t)]$
\end{center}
where
\begin{center}
$\displaystyle \hat{B}(s,t)= \frac{1}{t}\biggl[\ln\frac{d\sigma}{dt} -
\left.\ln\frac{d\sigma}{dt}\right|_{t=0}\biggr] \approx B(s,t) $
\end{center}
at small $t$. Both local slope and phase can be related to the
average transversal extent of the interaction region being the
averaging goes over elastic scattering events only\cite{Ku} :
\begin{center}
$\displaystyle \langle b^2\rangle_{\footnotesize\mbox{el}} =
\biggl\langle (-t)B^2(s,t)+ 4(-t)\biggl[\frac{\partial
\Phi(s,t)}{\partial t}\biggr]^2\biggr\rangle$
\end{center}
Unfortunately, practical application of such expressions is rather
model-dependent.  More tractable is the quantity $B(s)=
B(s,t=0)$\cite{Ku}:
\begin{equation}
\displaystyle 2B(s)= \langle b^2\rangle_{\footnotesize\mbox{tot}} -2\rho(s)\frac{\partial\Phi}{\partial t} (s,0)
\label{eq:3.1}
\end{equation}
where
\begin{center}
$\frac{\displaystyle \int db^2\,b^2\,{\rm
Im\,\tilde{T}}(s,b)}{\displaystyle \int db^2\,{\rm
Im\,\tilde{T}}(s,b)}=\langle b^2\rangle_{\footnotesize\mbox{tot}},$
\end{center}
and
\begin{center}
$\displaystyle \tilde{T}(s,b)= \frac{1}{16\pi s}\int
T(s,t)J_0(b\sqrt{-t})dt$
\end{center}
$\tilde{T}(s,b)$ is the scattering amplitude in the
impact parameter representation and the average is taken over all
possible processes in the given collision. We have to note that the derivative of the phase at $ t=0 $ also introduces a model-dependence, though in some models
this term is negligible.

 Finally, the diffraction peak width $\tilde{\Delta}$ is considered as the half-width
(starting from $t=0$) of the rectangle of the height
$\displaystyle \frac{d\sigma}{dt} (s,t=0)$, the area of which
gives $\sigma_{\footnotesize\mbox{el}}$. No direct relation to the average distances can be drawn.

\section*{4. General bounds} \label{sec:4}
There exists quite a trivial, but correct and useful, relation:
\begin{center}
$\langle B(s,t)(-t)\rangle = 1$
\end{center}
which can limit model expressions for $B$. If the simple
parametrization
\begin{equation}
\displaystyle
\frac{d\sigma}{dt}=\frac{d\sigma}{dt}(s,0)\exp[tB(s)]
\label{eq:4.1}
\end{equation}
were valid in all significant region of integration in $t$ then we
would have:
\begin{center}
$B(s,t) = B(s)= \tilde{\Delta}^{-1}$
\end{center}
but we know that's not the case. Then we can use the Heisenberg
uncertainty relations and get
\begin{center}
$\sqrt{\langle b^2
\rangle_{\footnotesize\mbox{el}}}\sqrt{\langle -t \rangle}\geq1$.
\end{center}
It seems that this inequality is the only source to estimate
$\langle b^2 \rangle_{\footnotesize\mbox{el}}$ in terms of an
observable quantity, $\langle -t \rangle$.

Let's now turn to the
bounds from general principles of quantum field theory. The upper
bound was derived \cite{roy} for the forward slope
\begin{center}
$\displaystyle B(s)\leq \frac{1}{8m_{\pi}^{2}} \ln^2
\frac{s}{s_{1}^{2}\sigma_{\footnotesize\mbox{tot}}} \equiv B_{\max}(s)$
\end{center}
at $s \gg s_{1}$. With $s_{1}=100~\mbox{GeV}^{2}$ and
$\sigma_{\footnotesize\mbox{tot}}(7~\mbox{TeV})=98.0\pm 2.5$ mb
\cite{Antchev-EPL-101-21004-2013} we get
\begin{center}
$B(s)\leq 56.8~\mbox{GeV}^{-2}$.
\end{center}
In compare with the reported values of $B(7~\mbox{TeV}) \simeq
20~\mbox{GeV}^{-2}$ the bound doesn't seem very restrictive,
though not awfully far. The unknown value of $ s_{1} $ introduces additional indeterminacy.

  In contrast, the lower bound does much
better. The bound \cite{macdowell}  reads (we neglect the values
of $\rho^{2}$ and $1/s$):
\begin{center}
$\displaystyle B(s)\geq
\frac{\sigma_{\footnotesize\mbox{tot}}^{2}}{18\pi
\sigma_{\footnotesize\mbox{el}}} \equiv B_{\min}(s)$.
\end{center}
With $\sigma_{\footnotesize\mbox{tot}}(7~\mbox{TeV}) \simeq
100~\mbox{mb}$ while $
\sigma_{\footnotesize\mbox{el}}(7~\mbox{TeV}) \simeq 25~\mbox{mb}$
the lower bound predicted from the general principles is
$17.7~\mbox{GeV}^{-2}$. Quite close to the data, indeed. If to use
the parametrization (\ref{eq:4.1}) then this lower bound could seem
quite trivial because, in this approximation,
\begin{center}
 $ \displaystyle B(s)\approx
\frac{\sigma_{\footnotesize\mbox{tot}}^{2}}{16\pi
\sigma_{\footnotesize\mbox{el}}} .$
 \end{center}
However, we cannot say that such a parametrization is 100\%
feasible. According to axiomatic QFT \cite{Cerulus-PL-8-80-1964}
it cannot be valid at all transferred momenta  and so higher
powers of $t$ in the exponent appear quite essential in derivation
of the slope from the (extrapolated) data. Accordingly, the
measurements by TOTEM \cite{TOTEM-NPB-899-527-2015} show that even
at smallest achieved $t$ the simple exponential does not do well and misses
some fine structures.

\section*{5. Elementary geometry of collision and the proton size from
the (low energy) $pp$ data} \label{sec:5}

Let us look again at the compilation of the data on the forward
slope (Fig. \ref{fig:1-2017}b). We observe that from the threshold energy $B(s)$ grows quite steeply (in logarithmic scale) till the energy region between $\simeq$ 10--11 GeV where it slows down. What happens in this energy interval? Is such a behaviour
expected from general considerations?

\begin{figure}[!h]
\begin{center}
\includegraphics[width=10.0cm,height=9.5cm]{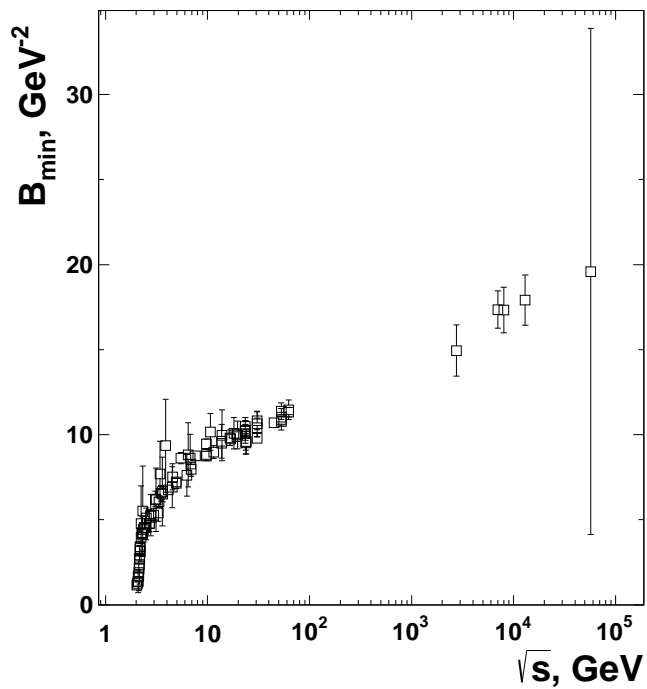}
\end{center}
\caption{\textit{The "data" for  $B_{\min}$ obtained with help of the DB17+ and results for $\sigma_{\footnotesize\mbox{el}}$ from \cite{Nemes-PoS-DIS2017-059,Antchev-arXiv-1712.06153,pdg-2016}. The estimation at $\sqrt{s}=57$ TeV is derived from corresponding measurements for total and inelastic cross-sections \cite{PAO-PRL-109-062002-2012}.}} \label{fig:2-2017}
\end{figure}

Let's take a simple example from the
quantum mechanical NR scattering via potential which is equivalent
to totally absorbing scatterer of radius $R$. It is well known
that relevant impact parameters are effectively cut-off: $b\leq
R$. The growing $\langle b^{2} \rangle$ would mean the growing
radius of the scatterer (absorber). QM has no answer, however, \textit{why}
should it grow. In optical language it could be expressed as the visible
size of a body would depend on the wavelength of light, quite a
strange phenomenon, indeed. If we take the lower bound shown above
we get the growing curve of the "data" points (see. Fig.
\ref{fig:2-2017}) but this in no way implies, only hints, in the best,
the "knee" we see in the behaviour of the slope itself.

So, our scarce information doesn't give us a physical insight to
understand the observed behaviour of the slope.

Let's tackle the
problem from another point. When we argue about such and such
impact parameters this actually mean that we deal with point like
particles in collision and so the impact parameter (averaged) is a
direct measure of the interaction field extent between them.
However, everybody knows that nucleons  are  in no case point like
 and show up their composite structure in many ways. Sure, one can
argue that they are practically point like if only one can neglect their sizes in comparison with the the interaction region radius. But can we? Let us try to find some relation between the sizes of nucleons and the slope.

\begin{figure}[!h]
\begin{center}
\includegraphics[width=10.0cm,height=10.0cm]{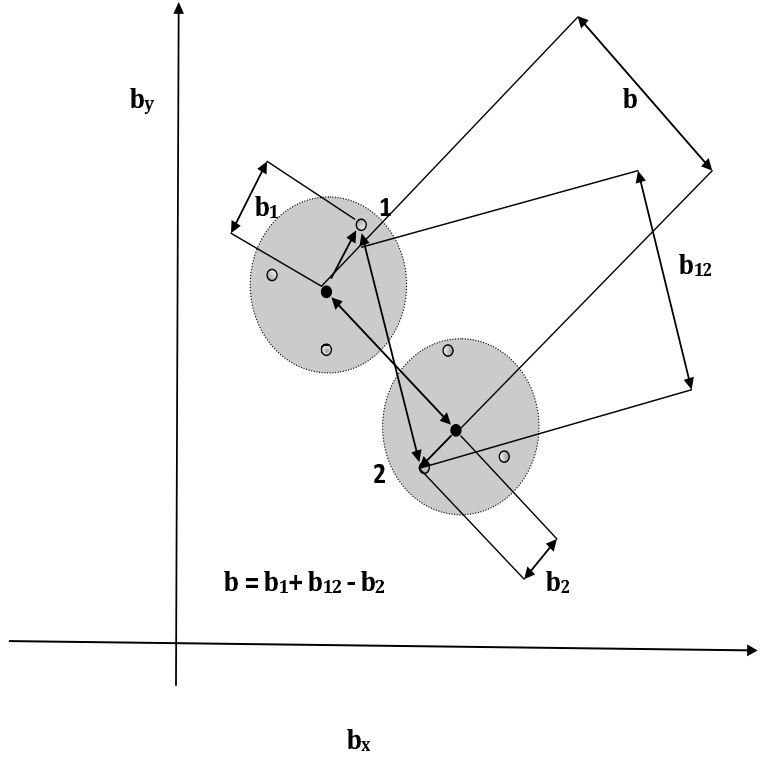}
\end{center}
\caption{\textit{Geometry of collision of extended particles.}}
\label{fig:3}
\end{figure}

Fig. \ref{fig:3} represents the "elementary geometry" of collision
of extended particles which will be  identified
with nucleons in what follows.

We have the following vector relation in $ \textbf{b}$-space
\begin{equation}
\textbf{b} = \textbf{b}_{1} + \textbf{b}_{12} - \textbf{b}_{2}.
\label{eq:5.1}
\end{equation}
Here $ \textbf{b} $ is the impact parameter of the colliding
nucleons, or the radius-vector between their "centres" in the
plane transverse to the collision axis while $ \textbf{b}_{1,2} $
denote the position of the interacting points of "strongly
interacting matter" in nucleons. At last, $ \textbf{b}_{12} $ is
the vector corresponding to strong interaction forces between the
sources inside the nucleons. Impact parameter \textit{per se} is
inaccessible for us in experiments, so we could only rely on some
average values. In the absence of polarization (which we assume
here) all the amplitudes in $\textbf{b}$-space are even under
$\textbf{b} \rightarrow  - \textbf{b}$ so we should take the
average of $\textbf{b}^{2}$ i.e. the following relation is valid\footnote{Here we follow the practice (not always justified) to omit the contribution from the phase [cf. Eq. (2)].} $2B = \bigl\langle
\textbf{b}^{2}\bigr\rangle$.  From Eq. (\ref{eq:5.1}) we get
\begin{equation}
\bigl\langle \textbf{b}^{2}\bigr\rangle = \bigl\langle
\textbf{b}_{1}^{2}\bigr\rangle + \bigl\langle
\textbf{b}_{2}^{2}\bigr\rangle + \bigl\langle
\textbf{b}_{12}^{2}\bigr\rangle - 2\bigl\langle
\textbf{b}_{1}\cdot\textbf{b}_{2}\bigr\rangle + \dots \label{eq:5.2}
\end{equation}
Here \dots means the rest of average correlations. According to
Eq. (\ref{eq:3.1}) the averaging is taken with the probability density
\begin{center}
$\displaystyle w(b^{2}) = \frac{{\rm
Im\,\tilde{T}}(s,b)}{\displaystyle \int db^2\,{\rm
Im\,\tilde{T}}(s,b)} $.
\end{center}

Sure, ${\rm Im\, \tilde{T}}(s,b)$ should, in a detailed
theory/model, contain explicit information about the internal
structure of colliding nucleons which provide averaging with the
strong interacting matter distributions in nucleons as well as the
averaging of the field strength extent with one or another
interaction mechanism. In the case if we know that nucleons do not
overlap by their proper extents given by $\bigl\langle
\textbf{b}^{2}_{1,2}\bigr\rangle $ the formula (\ref{eq:5.2})
simplifies to
\begin{center}
$\bigl\langle \textbf{b}^{2}\bigr\rangle = \bigl\langle
\textbf{b}_{1}^{2}\bigr\rangle + \bigl\langle
\textbf{b}_{2}^{2}\bigr\rangle + \bigl\langle
\textbf{b}_{12}^{2}\bigr\rangle$
\end{center}
We assume that quantities $\bigl\langle \textbf{b}^{2}_{1,2}\bigr\rangle $
are "genuine transverse sizes" independent of relativistic boosts , so the only source for the energy dependence of the forward slope is the term $\bigl\langle
\textbf{b}^{2}_{12}\bigr\rangle $.

 From this viewpoint we do not
see more natural explanation of the "knee" in the energy evolution
of $\bigl\langle \textbf{b}^{2} \bigr\rangle \approx 2B(s)$ as the onset
of the regime when colliding nucleons cease to overlap with each
other in the plane of impact parameter, i.e. when the average
extent achieves its minimum value in the absence of overlapping\footnote{Naive view of non-overlapping would be the condition $ \bigl\langle b^{2}\bigr\rangle^{1/2} \geq \bigl\langle b_{1}^{2}\bigr\rangle^{1/2} + \bigl\langle b_{2}^{2}\bigr\rangle^{1/2} $. However, this would mean a strong correlation $ \bigl\langle \textbf{b}_{1}\textbf{b}_{2}\bigr\rangle = \bigl\langle b_{1}^{2}\bigr\rangle^{1/2} \times \bigl\langle b_{2}^{2}\bigr\rangle^{1/2}  $ for which we do not see natural reasons.}
\begin{center}
$\bigl\langle \textbf{b}^{2}\bigr\rangle = \bigl\langle
\textbf{b}_{1}^{2}\bigr\rangle + \bigl\langle
\textbf{b}_{2}^{2}\bigr\rangle$
\end{center}
or
\begin{equation}
\displaystyle B(s) = \frac{1}{2}\bigl\langle
\textbf{b}^{2}\bigr\rangle = \frac{1}{2}\Bigl[\bigl\langle
\textbf{b}_{1}^{2}\bigr\rangle + \bigl\langle
\textbf{b}_{2}^{2}\bigr\rangle\Bigr] =  \bigl\langle
\textbf{b}^{2} \bigr\rangle_{N} \label{eq:5.3add}
\end{equation}
Here $\bigl\langle \textbf{b}^{2} \bigr\rangle_{N}$ means the
average square of the proper nucleon size as seen in the
transverse (impact parameter) plane being
\begin{equation}
\displaystyle \bigl\langle \textbf{b}^{2} \bigr\rangle_{N} =
\frac{2}{3}\bigl\langle r^{2}_{N} \bigr\rangle \label{eq:5.3}
\end{equation}
with $\textbf{r}$ meaning 3D radius. The change in the energy dependence of $B$ lies somewhere between 10 and 11 GeV . The average value of $B$ in this interval is \cite{Okorokov-AHEP-2015-914170-2015}
\begin{center}
$\langle B \rangle = 11.10 \pm 0.26~\mbox{GeV}^{-2}$.
\end{center}
From Eqs. (\ref{eq:5.3add}) and (\ref{eq:5.3}) we can estimate the nucleon radius as seen in $pp$-scattering:
\begin{center}
$\displaystyle \bigl\langle r^{2}_{N}\bigr\rangle = \frac{3}{2}\,\bigl\langle B \bigr\rangle = 16.65\pm 0.39 ~\mbox{GeV}^{-2} = (0.805 \pm 0.009~\mbox{fm})^{2}.$
\end{center}

We see that the nucleon size extracted from the $pp$-data at $\mathcal{O}(10~\mbox{GeV})$ "prefers" the CODATA  value indicated above in Sec. \ref{sec:1}.

\begin{figure}[!h]
\begin{center}
\includegraphics[width=14.0cm,height=9.0cm]{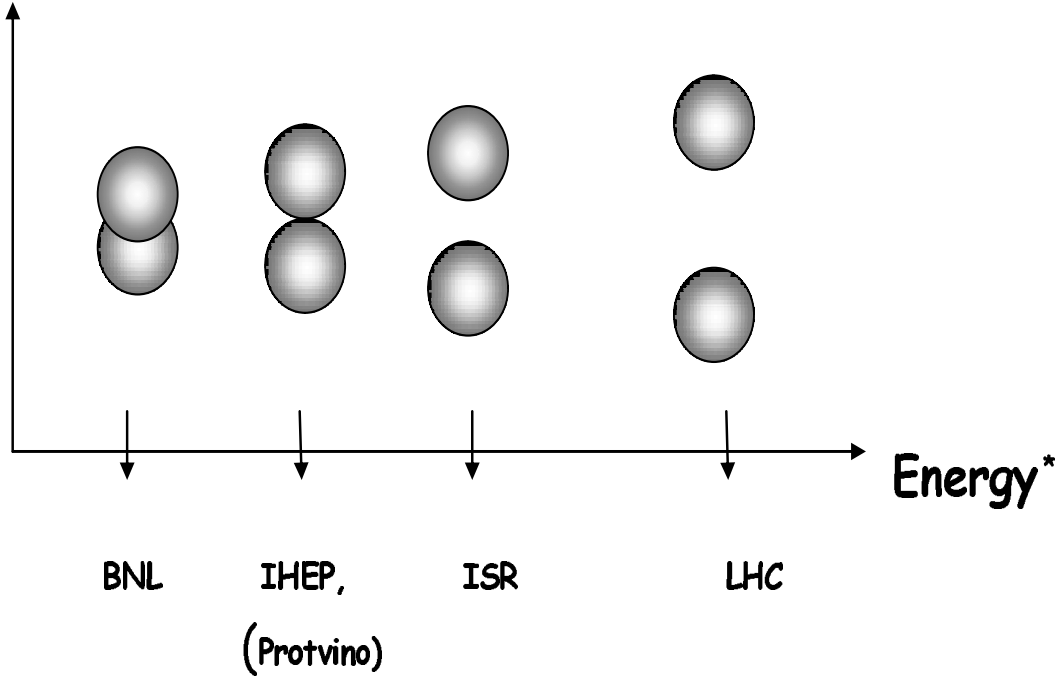}
\end{center}
\caption{\textit{Schematic view of collision at various energies.}}
\label{fig:5}
\end{figure}

Pictorially, the evolution of the $pp$ collision in the impact parameter plane looks as an extremely slow detaching of nucleon valence cores from each other (Fig. \ref{fig:5}).

\section*{6. Does the "effective" hadron size depend on the interaction type?} \label{sec:6}

We have seen that the proton size can be extracted from the data on the electromagnetic interaction. Have we right to use these sizes when discussing the strong interaction? It is in the spirit of quantum theory to investigate the influence of the measuring devices on the measurable quantities so the question is not idle.  To see the problem better, let us consider a simple model when the hadron $A$ is characterized by its valence quark distribution function $v_{A}(x, \textbf{b})$ \cite{Ki} where, as we already designated above, $x$ means the hadron momentum fraction carried away by the quark while $\textbf{b}$ is the transverse position of the quark counted from the "hadron center". The distribution has an evident normalization:
\begin{center}
$\displaystyle \int dxd^{2}b  v_{A}(x,\textbf{b})= N_{A}$,
\end{center}
where $N_{A}$ is the number of valence quarks. For the sake of simplicity  we will consider valence quarks of identical flavour. Let us assume that the interaction of hadrons can be described in the impulse approximation, i.e. when only one pair of quarks from colliding hadrons interact (with all possible pairs accounted and having in mind further account for multiple quark interactions in the eikonal framework). In the impact parameter representation such a "Born" amplitude, which we will take imaginary, $i \Omega_{AB}(s,\textbf{b})$, looks as follows:
\begin{equation}
\displaystyle \Omega_{AB}(s,\textbf{b}) =  \int dx_{1}d^{2}b_{1}dx_{2}d^{2}b_{2} v_{A}(x_{1},\textbf{b}_{1})v_{B}(x_{2},\textbf{b}_{2})\omega(sx_{1}x_{2}; \textbf{b}-\textbf{b}_{1}+\textbf{b}_{2}).
\end{equation}

Here $ \omega(s,\textbf{b})$ bears the meaning of the valence quark-quark  scattering amplitude with impact parameter $ \textbf{b} $ and c.m.s. energy $ \sqrt{s} $ .

If we assume that quark-quark interaction is local (in $\textbf{b}$-space):
\begin{center}
$\displaystyle \omega(sx_{1}x_{2}; \textbf{b}-\textbf{b}_{1}+\textbf{b}_{2})=K \delta(\textbf{b}-\textbf{b}_{1}+\textbf{b}_{2})$
\end{center}
then we rediscover the celebrated Chou--Yang formula \cite{Ch1} for the opacity:
\begin{center}
$\displaystyle\Omega_{AB}^{\footnotesize\mbox{Chou-Yang}}(s,\textbf{b})=K \int d^{2}b^{'}D_{A}(\textbf{b}-\textbf{b}^{'})D_{B}(\textbf{b}^{'})$
\end{center}
where "hadronic matter density" is
\begin{center}
$\displaystyle D_{A}(\textbf{b}) = \int {dx}v_{A}(x,\textbf{b})$.
\end{center}
In the momentum space with the 2D momentum transfer $\textbf{q}$ this looks as follows:
\begin{center}
$\displaystyle \hat{\Omega}_{AB}(s,\textbf{\textbf{q}}) =  \int \frac{dx_{1}}{x_{1}}\frac{dx_{2}}{x_{2}}\hat{v}_{A}(x_{1}, \textbf{q})\hat{v}_{B}(x_{2},\textbf{q})\hat{\omega}(sx_{1}x_{2};\textbf{q})$.
\end{center}
where $\textbf{q}$-space is related with $\textbf{b}$-space via 2D Fourier transform
\begin{eqnarray*}
\displaystyle \hat{\Omega}_{AB}(s,\textbf{\textbf{q}})=4s \int d^{2}b
\exp(i\textbf{qb})\Omega_{AB}(s,\textbf{b}),\\
\hat{v}_{A}(x,\textbf{q})=\int d^{2}b\exp(i\textbf{qb})v_{A}(x,\textbf{b})~~\qquad.
\end{eqnarray*}
Then we get that in the Born approximation the total $AB$-cross-section
is
\begin{center}
$\displaystyle \sigma_{\footnotesize\mbox{tot}}^{AB}(s)=\frac{\hat{\Omega}_{AB}(s,\textbf{\textbf{0}})}{s} = \frac{1}{s}\int \frac{dx_{1}}{x_{1}} \frac{dx_{2}}{x_{2}}v_{A}(x_{1})v_B(x_{2})\hat{\omega}(sx_{1}x_{2};\textbf{0})$
\end{center}
where $v_{A}(x)=\hat{v}_{A}(x_{1},\textbf{0})$ is the valence quark density in the momentum fraction which is familiar from the partonic analysis of DIS.
Let's first assume that the quark-quark interaction is mediated by a spin-$J$ boson:
\begin{center}
$\displaystyle \hat{\omega}(\hat{s};\textbf{q})=\hat{s}^{J}g^{2}\theta(\hat{s}-s_{0}) / (m^{2}(J)+\textbf{q}^{2})$.
\end{center}
In this case the Born cross-section is
\begin{center}
$\displaystyle \sigma_{\footnotesize\mbox{tot}}^{AB}(s)= \gamma_{J}\tilde{v}_{A}(J)\tilde{v}_{B}(J)(s/s_0)^{J-1}$
\end{center}
where $\displaystyle \tilde{v}_{A}(J)=\int_{0}^{1}dx x^{J-1}v_{A}(x)$ is the Mellin transform of the quark density and $\displaystyle \gamma_{J}=\frac{4g^{2}}{s_{0}m^{2} (J)}$.
We see that only for the vector exchange ($J=1$) we have the total cross-section proportional to the numbers of valence quarks in the colliding hadrons.
If, in the spirit of Van-Hove \cite{VH}, we sum up all possible exchanges in the $ t$-channel we get a Reggeized ($J \rightarrow \alpha(\textbf{q})$) version of the above said:
\begin{center}
$\displaystyle \sigma_{\footnotesize\mbox{tot}}^{AB}(s)= \gamma_{\alpha (0)}\tilde{v}_{A}(\alpha(0))\tilde{v}_{B}(\alpha(0))(s/s_0)^{\alpha(0)-1}.$
\end{center}
We see that the once celebrated  "quark counting rule" \cite{levin} holds only in the case of the "primordial" Pomeron with $ \alpha_{\mathbf{P}}(0)=1$. In this case the total cross-section(in the considered approximation) is proportional to the product of the valence quark numbers in the colliding hadrons:
\begin{center}
$\sigma_{\footnotesize\mbox{tot}}^{AB}(s)\rightarrow \mbox{const} N_{A}N_{B}.$
\end{center}

Let us come back to the effective sizes of the colliding hadrons. According to Eq. (8) we obtain for the size of the transverse interaction region $\bigl\langle b^{2}\bigr\rangle_{AB}$ :
\begin{equation}
\displaystyle \bigl\langle b^{2}\bigr\rangle _{AB} = \bigl\langle {b}^{2} \bigr\rangle_{A}(\Delta)+\bigl\langle {b}^{2} \bigr\rangle_{B}(\Delta)+
4\alpha^{'}_{\mathbf{P}}\ln(s/s_{\footnotesize\mbox{eff}}),~~~ \Delta \equiv \alpha_{\mathbf{P}}(0)-1. \label{eq:7.1}
\end{equation}
This formula is essential. First of all we see that "effective sizes" of the hadrons A and B
\begin{center}
$\displaystyle \bigl\langle {b}^{2} \bigr\rangle_{A,B}(\Delta) \doteq \biggl[\int dx v_{A,B}(x)x^{\Delta}\rho^{2}_{A,B}(x)\biggr]\biggl[\displaystyle \int dx v_{A,B}(x)x^{\Delta}\biggr]^{-1}$,
\end{center}
where
\begin{center}
$\displaystyle \rho^{2}_{A,B}(x) = \biggl[\int d^{2}b b^{2}v_{A,B}(x,\textbf{b})\biggr]\biggl[\int d^{2 }b v_{A,B}(x,\textbf{b})\biggr]^{-1}= 4\left.\frac{\partial[\ln\hat{v}_{A,B}(x,\textbf{q})]}{\partial t}\right|_{t=0} $
\end{center}
generally may significantly differ (for the "supercritical" Pomeron with $\Delta > 0$) from their "natural sizes" extracted from the electro-magnetic form factors. The latter have the following form in terms of quark densities:
\begin{center}
$\displaystyle \bigl\langle {b}^{2} \bigr\rangle_{A,B} \doteq \biggl[\int dxv_{A,B}(x)\rho^{2}_{A, B}(x)\biggr]\biggl[\int dx v_{A, B}(x)\biggr]^{-1}$.
\end{center}
We also observe that the energy dependence of the interaction region is influenced by the fact that the driving interaction is provided by quarks which carry lower energy than the colliding hadrons. Indeed, instead of the energy scale factor $s_{0}$ we get now the larger "effective threshold"
\begin{center}
$s_{\footnotesize\mbox{eff}}=s_{0}\exp\bigl[-\langle\ln x\rangle_{A}-\langle\ln x\rangle_{B}\bigr]> s_{0}$
\end{center}
where
\begin{equation}
\displaystyle \langle\ln x\rangle_{A,B} = \biggl[\int_{0}^{1} dx x^{\Delta} v_{A,B}(x)\ln x\biggr]\biggl[\int_{0}^{1} dx x^{\Delta}v_{A,B}(x)\biggr]^{-1}. \label{eq:7.2}
\end{equation}

In modelling practice the values of $\Delta$ vary  dependent on the model in question. From such a general form as Eq. (\ref{eq:7.2}) we can't say much.
So, just to get an idea of the influence of the non-zero $\Delta$ let us consider a "toy" model for the valence quark densities which roughly bear general properties of those observed in DIS but are designed in a simplest form to illustrate things. We assume the valence quark density of the 3-quark nucleon in the form
\begin{center}
$\displaystyle v_{A}(x) = \frac{3}{2\sqrt{x}}\vartheta(1-x)$
\end{center}
which obeys the sum rules
\begin{center}
$\displaystyle \int_{0}^{1}dx v_{A}(x)=N_{A}=3$,
\end{center}
and
\begin{center}
$\displaystyle \int_{0}^{1}dxx v_{A}(x)=1$.
\end{center}
It has the secondary Regge behaviour $\sim x^{-\alpha_{\mathbf{R}}(0)}$ at small $x$ (we assume $\alpha_{\mathbf{R}}(0)=1/2$), though is not trustworthy at $x \rightarrow 1$. The "sea" partons are assumed to be absorbed by the valence quarks. More detailed formula with account of the impact parameter is as follows:
\begin{center}
$\displaystyle  v_{A}(x,\textbf{b})= \frac{v_{A}(x)}{\pi \rho_{A}^{2}(x)}\exp\bigl[-b^{2}/\rho_{A}^{2}(x)\bigr]$
\end{center}
where
\begin{center}
$\displaystyle \rho_{A}^{2}(x)= \rho_{A}^{2}(\mbox{core})-4\alpha_{\mathbf{R}}^{'}(0)\ln x$.
\end{center}
Thus, we get
\begin{center}
$\displaystyle \bigl\langle b^{2} \bigr\rangle_{A}=\rho_{A}^{2}(\mbox{core})+ 4\alpha_{\mathbf{R}}^{'}(0)/[1-\alpha_{\mathbf{R}}(0)]$.
\end{center}
The notation "core" means in this context the region where reggeons are emitted from. Comparing this with Eq. (\ref{eq:5.3add}), numerical value for $\langle B\rangle$ from Sec. \ref{sec:5} and assuming the generic value $\alpha_{\mathbf{R}}^{'}(0)\approx 1~\mbox{GeV}^{-2}$ we get $\rho_{A}^{2}(\mbox{core})\approx 3~\mbox{GeV}^{-2}$ i.e. the sources of the virtual $\rho$-, $\omega$-, $f$-mesons reside at the "core" of the size $\approx 0.35$ fm. This size is somewhat intermediate between the core size (0.2 fm) and that of the "baryon number shell" (0.44 fm) as argued in \cite{islam}.

Let us now estimate to which extent the effective radii of colliding hadrons and the effective energy scale in the slope evolution are being changed  when passing from the photon to the Pomeron exchange.
Specifically,
\begin{equation}
\displaystyle \bigl\langle {b}^{2} \bigr\rangle_{A}(\Delta) = \rho_{A}^{2}(\mbox{core)}+ 8\alpha_{\mathbf{R}}^{'}(0)/[1+2\Delta]. \label{eq:7.3}
\end{equation}

So, for $\Delta = 0.05$ we get
\begin{center}
$\bigl[\bigl\langle b^{2} \bigr\rangle_{N,\,\,\footnotesize\mbox{eff}}\bigr]^{1/2} = 0.64~\mbox{fm}$
\end{center}
and for $\Delta = 0.2$ we get
\begin{center}
$\bigl[\bigl\langle b^{2} \bigr\rangle_{N,\,\,\footnotesize\mbox{eff}}\bigr]^{1/2} = 0.59~\mbox{fm}$
\end{center}
to compare with the "genuine" size $\approx 0.66$ fm from Sec. \ref{sec:5}. We see that for the intercepts in the considered range the effective sizes of the nucleons change insignificantly. In contrast,the change of the effective energy scale may be more noticeable. In fact,
\begin{equation}
\displaystyle s_{\footnotesize\mbox{eff}} = s_{0}\exp\bigl[-\langle \ln x\rangle_{A}(\Delta)-\langle \ln x\rangle_{B}(\Delta)\bigr] = s_{0}\exp\bigl[4/(1+2\Delta)\bigr]. \label{eq:7.4}
\end{equation}
At $\Delta = 0.1$ we get $s_{\footnotesize\mbox{eff}} = 28s_{0}$. It is amusing that, in this case, if we take $s_{0}=(2m_{N}+ m_{\pi})^2$, the lowest inelastic threshold of the nucleon-nucleon collision, then
\begin{center}
$s_{\footnotesize\mbox{eff}} = 113.8~\mbox{GeV}^{2}$ or $\sqrt{s_{\footnotesize\mbox{eff}}} = 10.7 ~\mbox{GeV}$.
\end{center}
We are caught exactly in the region of energies where the forward slope slows down its growth while the total cross section begins to increase(see Sec. \ref{sec:5}). So these, relatively coarse, estimates show that \textit{grosso modo} the reasoning presented in  Sec. \ref{sec:1} looks very plausible.

\section*{7. Futuristic prognosis for the slope and deceptive an\-ti\-ci\-pa\-ti\-ons of the "truly asymptotic regime"} \label{sec:7}

Let us come back to the forward slope evolution. From the preceding Section we have learned that the energy dependence of the interaction region caused by the Pomeron exchange begins to reveal itself only at relatively high energies of the order of $\approx 10$ GeV in the c.m.s. Actually the effective sizes of the colliding hadrons comprise the lion's share of the interaction region, till the LHC energies. Composite nature of colliding hadrons stipulates the slow evolution of the forward slope due to minimizing  the effective energies of collision provided by valence quarks for average quantities like the interaction region size. For bare cross-sections this circumstance acts in the opposite direction enforcing the early necessity in "unitarization". This is a hand-wave explanation of the significant disparity in energy dependence between the total cross-section and the slope.

For the sake of concreteness let us again use a simple Regge-eikonal model where the eikonal is to be given by the function $\Omega_{AB}(s,\textbf{b})$ considered in the previous Section. It is easy to derive the following expression for the average impact parameter in this framework (for brevity we will omit indices $A$, $B$ specifying colliding hadrons):
\begin{equation}
\bigl\langle b^{2}\bigr\rangle = \kappa(s) \bigl\langle b^{2}\bigr\rangle^{1\mathbf{P}}
\end{equation}
where $\bigl\langle b^{2}\bigr\rangle^{1\mathbf{P}}$ defined by Eq. (\ref{eq:7.1}) is the one-Pomeron approximation for the transverse interaction radius and the coefficient $\kappa(s)$ provides the account of  the s-channel unitarity:
\begin{equation}
\displaystyle \kappa(s) = \biggl[\sum_{k=1}^{\infty} \frac{1}{k^{2}k!}(-2\Omega(s,\textbf{0}))^{k}\biggr]\biggl[\sum_{k=1}^{\infty} \frac{1}{k k!}(-2\Omega(s,\textbf{0}))^{k}\biggr]^{-1} \label{eq:8.1}
\end{equation}
We have argued in preceding Sections that our arguments concerning the role of the proper nucleon size are valid starting from the "effective threshold" $\sqrt{s_{\footnotesize\mbox{eff}}}\simeq \mathcal{O}$(10 GeV) when inter-quark spatial correlations are neglected.

Despite the fact that the above formalism is of a general nature, the discussion below for possible signatures of asymptotic regime is focused precisely on proton-proton scattering because the experimental data up to the $\sqrt{s} \approx 100$ TeV are available for some scattering parameter from the set $\mathcal{G}_{pp}$ for $pp$ collisions only (Fig. \ref{fig:1-2017}a). Now, let us try to see to which extent the one-Pomeron expression for the slope
\begin{equation}
\displaystyle B^{1\mathbf{P}}_{pp}(s) = \bigl\langle b^{2} \bigr\rangle_{pp}^{1\mathbf{P}} / 2 = 2\alpha^{'}_{\mathbf{P}}(0)\ln(s/s_{\footnotesize\mbox{eff}}) + \bigl\langle b^{2} \bigr\rangle_{N,\,\,\footnotesize\mbox{eff}}  \label{eq:8.2}
\end{equation}
can describe the existing data. According to Eqs. (\ref{eq:7.3}), (\ref{eq:7.4})
\begin{center}
$\displaystyle s_{\footnotesize\mbox{eff}}= s_{0}\exp\biggl[\frac{2}{\alpha_{\mathbf{P}}(0)-\alpha_{\mathbf{R}}(0)}\biggr],~~ \bigr\langle b^{2} \bigr\rangle_{N,\,\,\footnotesize\mbox{eff}} = \rho_{p}^{2}(\mbox{core}) + \frac{4\alpha^{'}_{\mathbf{R}}(0)}{1+\Delta-\alpha_{\mathbf{R}}(0)}$
\end{center}
and $\displaystyle \bigl\langle b^2 \bigr\rangle_{pp}^{1\mathbf{P}} = 4\alpha^{'}_{\mathbf{P}}(0)\ln (s/s_{0})
+ 8\frac{\alpha_{\mathbf{R}}^{'}(0)-\alpha_{\mathbf{P}}^{'}(0)}{\alpha_{\mathbf{P}}(0)
-\alpha_{\mathbf{R}}(0)}+2\rho^{2}_{p}(\mbox{core})$. Finally, the energy dependence of the forward slope within one-Pomeron approach is following:
\begin{equation}
\displaystyle B^{1\mathbf{P}}_{pp}(s) = 2\alpha^{'}_{\mathbf{P}}(0)\ln (s/s_{0})
+ \biggl[4\frac{\alpha_{\mathbf{R}}^{'}(0)-\alpha_{\mathbf{P}}^{'}(0)}{\alpha_{\mathbf{P}}(0)
-\alpha_{\mathbf{R}}(0)}+\rho^{2}_{p}(\mbox{core})\biggr]. \label{eq:8.2.dop}
\end{equation}

\begin{figure}[!h]
\begin{center}
\includegraphics[width=10.0cm,height=9.5cm]{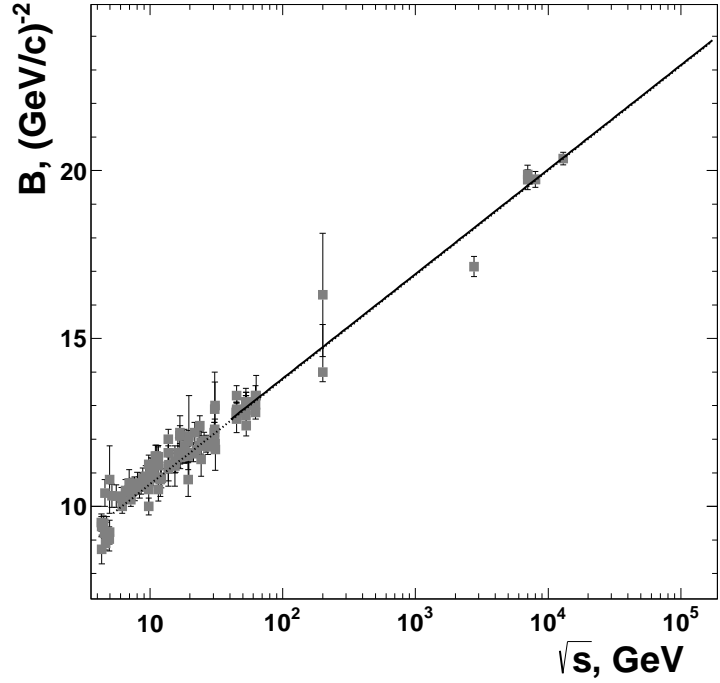}
\end{center}
\caption{\textit{Fit results for $B(s)$ in $pp$ scattering within one-Pomeron approach (\ref{eq:8.2.dop}) at two free parameters. Experimental data are from DB17+, the dashed curve is the fit in energy domain $\sqrt{s} \geq 5$ GeV, solid line -- for $\sqrt{s} \geq 40$ GeV.}} \label{fig:4add-2017}
\end{figure}

Let's take the values of the core radius, secondary Reggeon slope and intercept fixed:
\begin{center}
$\rho_{p}^{2}(\mbox{core})=3~\mbox{GeV}^{-2},~~\alpha_{\mathbf{R}}(0)=0.5,~~
\alpha^{'}_{\mathbf{R}}(0)= 1~\mbox{GeV}^{-2}$
\end{center}
while the values of $\Delta$, $\alpha^{'}_{\mathbf{P}}(0)$ remain adjustable parameters and we also put in this section $s_{0}=1$ GeV$^{2}$. It is shown \cite{Okorokov-AHEP-2015-914170-2015} that the function $\propto \ln s$ describes the experimental data for slope reasonably at $\sqrt{s} \geq 5$ GeV. Therefore,  $B(s)$ is fitted by (\ref{eq:8.2.dop}) at the lower boundary $\sqrt{s_{\min}}=5$ GeV. At the first stage we use Eq. (\ref{eq:8.2.dop}) with 5 free parameters: $\alpha^{'}_{\mathbf{P}}(0)$, $\Delta$, $\rho_{p}(\mbox{core})$, $\alpha_{\mathbf{R}}(0)$, $\alpha^{'}_{\mathbf{R}}(0)$ and it provides the results shown in Table \ref{tab:8-1add}. As seen, the fit values of the $\rho_{p}(\mbox{core})$, $\alpha_{\mathbf{R}}(0)$ and $\alpha^{'}_{\mathbf{R}}(0)$ agree with estimations assigned above quite reasonably. Then the energy dependence of the experimental slope is approximated by (\ref{eq:8.2.dop}) with 2 free parameters and fixed values of  $\rho_{p}(\mbox{core})$, $\alpha_{\mathbf{R}}(0)$, $\alpha^{'}_{\mathbf{R}}(0)$ in various energy ranges. Fit results are shown in Table \ref{tab:8-1add} and in Fig. \ref{fig:4add-2017}. Values of $\alpha^{'}_{\mathbf{P}}(0)$ and $\Delta$ are independent from $\sqrt{s_{\min}}$ and the one-Pomeron approximation(\ref{eq:8.2.dop}) describes the experimental data with statistically reasonable quality though the value of  $\alpha^{'}_{\mathbf{P}}(0)$ is significantly larger than the "nominal" value 0.25 GeV$^{-2}$. We see that the one-Pomeron expression agrees quite well with the LHC data at $\sqrt{s}=7$, 8 and 13 TeV but goes above the point at $\sqrt{s}=2.76$ TeV though not too much\footnote{As was noticed in \cite{Pietr}, the value of the slope given by the TOTEM Collaboration at this energy could signal (if definitely confirmed) the onset of some new regime of the interaction radius energy evolution.} (Fig. \ref{fig:4add-2017}).

\begin{table*}[!h]
\caption{\label{tab:8-1add}Values for fit parameters for approximation of the slope by function (\ref{eq:8.2.dop}).}
\begin{center}
\begin{tabular}{ccccccc}
\hline \multicolumn{1}{c}{$\sqrt{s_{\min}}$,} &
\multicolumn{5}{c}{Fit parameters } &
\multicolumn{1}{c}{$\chi^{2}/\mbox{ndf}$} \\\cline{2-6}
GeV& $\alpha^{'}_{\mathbf{P}}(0)$,& $\Delta \times 10^{2}$ & $\rho_{p}(\mbox{core})$,& $\alpha_{\mathbf{R}}(0) \times 10^{2}$ & $\alpha^{'}_{\mathbf{R}}(0)$& \\
 & GeV$^{-2}$& & GeV$^{-1}$ &  & GeV$^{-2}$ & \\
\hline
5  & $0.338\pm0.002$ & $9.40\pm0.07$ & $1.748\pm0.006$    & $44.20\pm0.12$ & $1.070\pm0.003$ & 1.43 \\
   & $0.338\pm0.002$ & $8.2\pm0.4$   & $\sqrt{3}$ (fixed) & $50$ (fixed)     & $1.0$ (fixed) & 1.39 \\
30 & $0.340\pm0.003$ & $8.40\pm0.10$ & --//-- & --//-- & --//-- & 1.12 \\
40 & $0.338\pm0.004$ & $7.7\pm1.1$   & --//-- & --//-- & --//-- & 1.30 \\
\hline
\end{tabular}
\end{center}
\end{table*}

If to use the Regge-eikonal formula \cite{prok} (we neglect the secondary Reggeon exchanges)
\begin{equation}
\sigma^{pp}_{\footnotesize\mbox{tot}}(s) = 2\pi \bigl\langle b^2
\bigr\rangle_{pp}^{1\mathbf{P}}[\mathbf{C}+\ln \xi-\mbox{Ei}(-\xi)]
\label{eq:8.3}
\end{equation}
with Euler's constant $\mathbf{C} = 0.5772...$ and $\displaystyle \xi(s) = 2\Omega_{pp}(s,\textbf{0})=
\frac{g^{2}(s/s_{0})^\Delta}{2\pi s_{0}\bigl\langle b^{2}\bigr\rangle_{pp}^{1\mathbf{P}}} > 0$, then the formulas for both  the total cross-section and the slope may be written in a more concise form (for brevity we omit index $pp$ in the left hand side):

\begin{center}
$\sigma_{\footnotesize\mbox{tot}}(s)= \sigma^{1\mathbf{P}}_{\footnotesize\mbox{tot},pp}(s)k(s),~~~~~\sigma^{1\mathbf{P}}_{\footnotesize\mbox{tot},pp}(s)= g^{2}(s/s_{0})^{\Delta}s_{0}^{-1}$
\end{center}
and
\begin{center}
$B(s)= B^{1\mathbf{P}}_{pp}(s)\kappa(s)$
\end{center}
which relate the measured cross-section and the slope with their "bare" (1-Pomeron) values   by "dressing factors" $k(s)$ and $\kappa(s)$ [defined by Eq. (\ref{eq:8.1})] which actually depend on an "effective evolution parameter" $\xi$:
\begin{equation}
\displaystyle k(s) = \frac{\mathbf{C}+\ln \xi-\mbox{Ei}(-\xi)}{\xi},~~~~~\kappa(s) =\xi\:
\frac{{}_{3}F_{3}(1,1,1;2,2,2;-\xi)}{\mathbf{C}+\ln
\xi-\mbox{Ei}(-\xi)},\label{eq:8.3.dop1}
\end{equation}
where
${}_{p}F_{q}(a_{1},\ldots,a_{p};b_{1},\ldots,b_{q};x)$ is the
generalized hypergeometric function. One can absorb
$s_{\footnotesize\mbox{eff}}$ into $\bigr\langle b^{2} \bigr\rangle_{N,\,\,\footnotesize\mbox{eff}}$ to make a new parameter $\displaystyle r_{0}^{2} \equiv 4\frac{\alpha_{\mathbf{R}}^{'}(0)-\alpha_{\mathbf{P}}^{'}(0)}{\alpha_{\mathbf{P}}(0)
-\alpha_{\mathbf{R}}(0)}+\rho^{2}_{p}(\mbox{core})$ and
rewrite

\begin{center}
$\displaystyle
\xi(s)=\frac{g^{2}}{4\pi}\frac{(s/s_{0})^\Delta}{s_{0}\bigl[r_{0}^{2}+2\alpha^{'}_{\mathbf{P}}(0)\ln
(s/s_{0})\bigr]}$.
\end{center}

Then the energy dependence of the forward slope is
\begin{equation}
B(s)=\bigl[r_{0}^{2}+2\alpha^{'}_{\mathbf{P}}(0)\ln
(s/s_{0})\bigr]\xi\:
\frac{{}_{3}F_{3}(1,1,1;2,2,2;-\xi)}{\mathbf{C}+\ln
\xi-\mbox{Ei}(-\xi)}. \label{eq:8.3.dop2}
\end{equation}
Behavior of $\xi(s)$ depends on \emph{a priori} unknown values of the
parameters $g$, $\Delta$, $r_{0}$ and
$\alpha^{'}_{\mathbf{P}}(0)$.
The following parameter ranges:
\begin{center}
$g=7.8 \pm
0.5$, $\Delta=0.095 \pm0.010$, $r_{0}=(2.8 \pm 0.3)$ GeV$^{-1}$
and $\alpha^{'}_{\mathbf{P}}(0)=(0.23 \pm 0.02)$ GeV$^{-2}$
\end{center}
has been chosen from some general assumptions. For these parameters
the function $\xi(s)$ (which is a much more relevant evolution parameter that just the collision energy, $\sqrt{s}$) is evaluated qualitatively for an energy domain from
$4m_{p}^{2}$ and up to the Plank scale $s_{\footnotesize\mbox{Pl}}$.
Correspondingly, the range of $\xi$ extends from $\sim 0.5$ up to the value $\sim 500$ and
the uncertainty of $\Delta$ dominates the spread of values of
$\xi$ at fixed $s$. Function $\xi(s)$ increases smoothly with
damping of the growth at $\xi \gg 1$. Hereby $\kappa(\xi) \approx$
1.0--1.1 at $\xi$ up to  1.0 which corresponds to
$\sqrt{s} < 140$ GeV and $\kappa(\xi)$ reaches 3.5--3.8 at
ultimate energies for parameter values under discussion. Detailed
analysis shows that $\kappa\bigl[\xi(s)\bigr]$ can be approximated by
functions
\begin{subequations}
\begin{equation}
\kappa(\xi) \approx f_{1}(\xi)=1+0.109\,\xi,~~~~~~~~~ \xi \leq
\xi_{1},\label{eq:8.3dop3a}
\end{equation}
\begin{equation}
\kappa(\xi) \approx f_{2}(\xi)=0.60+0.47\ln \xi,~~~ \xi \geq
\xi_{2},\label{eq:8.3dop3b}
\end{equation}\label{eq:8.3dop}
\end{subequations}
\hspace*{-0.2cm}quite well (the accuracy is better than
99\%), where $\xi_{1}=3$ and $\xi_{2}=16$ which correspond to
$\sqrt{s_{1}} \approx 0.57$ PeV and $\sqrt{s_{2}} \approx 3.3
\times 10^{10}$ GeV for median values of the parameters $g$,
$\Delta$, $r_{0}$ and $\alpha^{'}_{\mathbf{P}}(0)$ shown above.
Therefore, the approximating function $f_{1}(\xi)$ is usable within
the total energy range available  both for the present accelerator
experiments and the cosmic ray measurements as well as in any
future collider projects. The low energy boundary for
applicability of (\ref{eq:8.3dop3b}) approaches  the GUT
domain in order of magnitude. This allows us to use the following
approximation
\begin{equation}
\displaystyle B(s) \approx
r_{0}^{2}+2\alpha^{'}_{\mathbf{P}}(0)\ln (s/s_{0})+
0.109\frac{g^{2}(s/s_{0})^{\Delta}}{4\pi s_{0}} \label{eq:8.4}
\end{equation}
for experimentally available energy range.

We also consider
expressions of quite a popular form:
\begin{subequations}
\begin{equation}
\sigma_{\footnotesize\mbox{tot}}^{pp}=\sigma_{0}+2\alpha^{'}_{\mathbf{P}}(0)\ln(s/s_{0})+c_{2}\ln^{2}(s/s_{0}),\label{eq:8.5a}
\end{equation}
\begin{equation}
B(s)= b_{0}+2\alpha^{'}_{\mathbf{P}}(0)\ln(s/s_{0})+b_{2}\ln^{2}(s/s_{0}) \label{eq:8.5b}
\end{equation}\label{eq:8.5}
\end{subequations}
\hspace*{-0.2cm}which are similar to those suggested and used in
\cite{Schegelsky-PRD-85-094024-2012} as expressions which
allegedly account for multi-Pomeron exchanges quantified by the presence of "Froissart-like" terms $\sim \ln^{2}(s/s_{0})$.  Probably, to emphasize the closeness to the "true asymptotic regime" the authors of \cite{Schegelsky-PRD-85-094024-2012} have diligently chosen the ration $ c_{2}/b_{2} $ equal to the "due" value, $ 8\pi $.

As seen from  Eq. (\ref{eq:8.3dop3b}),  $\kappa(\xi)\propto \ln \xi$
only at $\xi \gg 1$. It means that a $\ln^{2}(s/s_{0})$--type asymptotic behavior both for $B(s)$ and
$\sigma_{\footnotesize\mbox{tot}}(s)$ is being achieved only at
$\xi \gg 1$, i.e. in an energy domain which lies far outside the LHC energies.

It is interesting to note that the approximation
$\kappa(\xi) = 1+0.109\xi$, we used above, is saturated in the
region of its applicability by 5--6 exchanged Pomerons.

\begin{figure}[!h]
\begin{center}
\includegraphics[width=12.0cm,height=13.0cm]{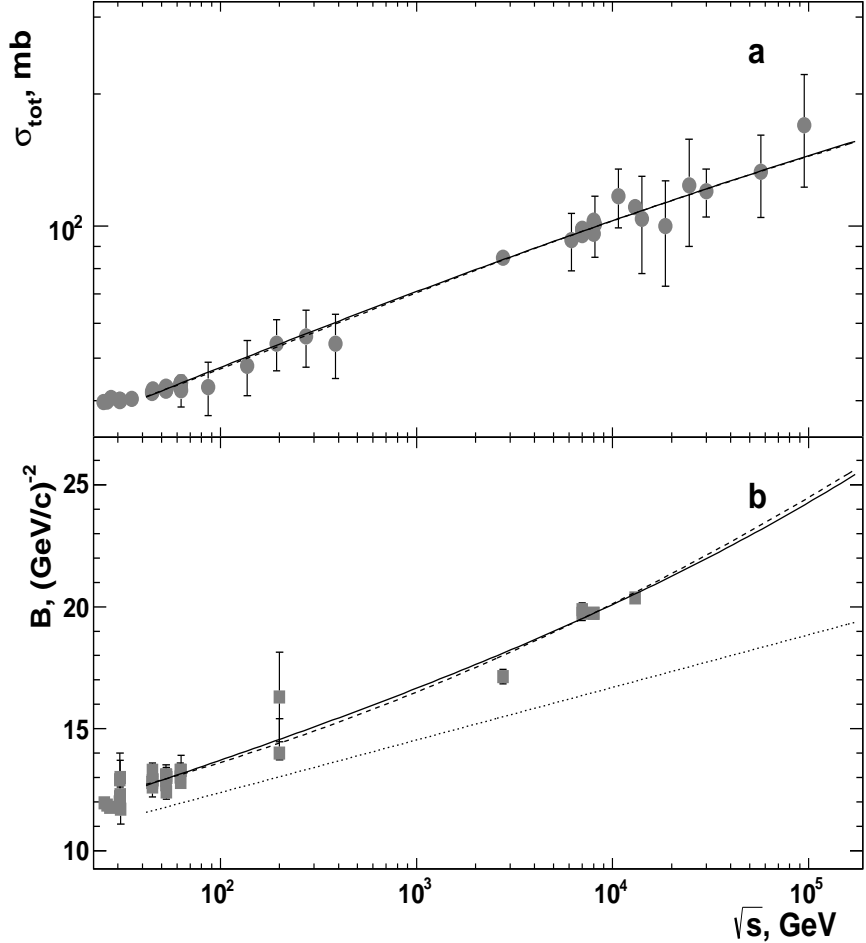}
\end{center}
\caption{\textit{Simultaneous fit results for $\sigma_{\footnotesize\mbox{tot}}$ (a) and slope (b) in $pp$ scattering in the  energy domain $\sqrt{s} \geq 40$ GeV. The solid curves correspond to the approximations (\ref{eq:8.3}), (\ref{eq:8.4}) within Regge-eikonal model and dashed curves are for Eqs. (\ref{eq:8.5}). In the bottom panel the dotted line is calculated within one-Pomeron approach (\ref{eq:8.2.dop}) with parameters obtained by simultaneous fit with general Regge-eikonal Eqs. (\ref{eq:8.3}), (\ref{eq:8.4}).}} \label{fig:4-2017}
\end{figure}

Let us now see how the above mentioned approximate expression for $B(s)$ works.
At Fig. \ref{fig:4-2017} the results of the simultaneous fitting of the total cross-sections
and the slopes according to both sets of approximations (\ref{eq:8.3}),
(\ref{eq:8.4}) and (\ref{eq:8.5}) are shown. The fits are made in the energy domains
$s \geq s_{\min}$ at various lower boundaries $s_{\min}$. Numerical values for fit parameters
are shown in Tables \ref{tab:8-1} and \ref{tab:8-2} for Regge-eikonal model and
approximations (\ref{eq:8.5}), respectively. Furthermore, in the first case the additional way is considered for simultaneous fit for the set of the scattering parameters $\mathcal{G}_{pp}$ with the one-Pomeron formula (\ref{eq:8.2.dop}) instead of Eq. (\ref{eq:8.4}). Thus in Table \ref{tab:8-1} the first line corresponds to the general Regge-eikonal relations and the second one to Eq. (\ref{eq:8.2.dop}) for $B(s)$, i.e. with $\kappa(s) \equiv 1$ for each $s_{\min}$.

\begin{table*}[!h]
\caption{\label{tab:8-1}Values for fit parameters for approximation of the set $\mathcal{G}_{pp}$ by various ways within Regge-eikonal model.}
\begin{center}
\begin{tabular}{cccccc}
\hline \multicolumn{1}{c}{$\sqrt{s_{\min}}$,} &
\multicolumn{4}{c}{Fit parameters} &
\multicolumn{1}{c}{$\chi^{2}/\mbox{ndf}$} \\\cline{2-5}
GeV& $r_{0}$, GeV$^{-1}$ &  $\alpha^{'}_{\mathbf{P}}(0)$, GeV$^{-2}$ & g & $\Delta \times 10^{2}$ & \\
\hline
30 & $2.782\pm0.019$ & $0.250\pm0.004$ & $8.09\pm0.05$ & $9.30\pm0.17$ & 2.04 \\
   & $2.743\pm0.017$ & $0.340\pm0.003$ & $8.11\pm0.05$ & $8.93\pm0.15$ & 2.01 \\
40 & $2.84\pm0.02$   & $0.234\pm0.005$ & $7.74\pm0.06$ & $10.2\pm0.2$  & 0.98 \\
   & $2.75\pm0.02$   & $0.338\pm0.004$ & $7.82\pm0.05$ & $9.65\pm0.16$ & 1.04 \\
\hline
\end{tabular}
\end{center}
\end{table*}
\begin{table*}[!h]
\caption{\label{tab:8-2}Values for fit parameters for approximation of the set $\mathcal{G}_{pp}$ by functions (\ref{eq:8.5}).}
\begin{center}
\begin{tabular}{ccccccc}
\hline \multicolumn{1}{c}{$\sqrt{s_{\min}}$,} &
\multicolumn{5}{c}{Fit parameters, GeV$^{-2}$} &
\multicolumn{1}{c}{$\chi^{2}/\mbox{ndf}$} \\\cline{2-6}
GeV& $\sigma_{0}$ & $\alpha^{'}_{\mathbf{P}}(0)$ & $c_{2}$ & $b_{0}$ & $b_{2}\times10^{2}$ & \\
\hline
30 & $75.8\pm0.6$ & $0.1132\pm0.0003$ & $0.515\pm0.010$ & $9.94\pm0.06$ & $1.77\pm0.03$ & 1.41 \\
40 & $72.6\pm0.8$ & $0.1129\pm0.0003$ & $0.552\pm0.011$ & $10.06\pm0.07$ & $1.74\pm0.03$& 0.86 \\
\hline
\end{tabular}
\end{center}
\end{table*}

Values of $g$ coincide within errors for Regge-eikonal simultaneous fits with (\ref{eq:8.2.dop}) and (\ref{eq:8.4}) while the corresponding values for $r_{0}$, $\Delta$ agree with each other within 2 s.d. at certain $\sqrt{s_{\min}}$. Two different versions of simultaneous fit for $\mathcal{G}_{pp}$ within Regge-eikonal model show the close qualities and corresponding curves are almost the same. But it should be noted that the simultaneous fit with (\ref{eq:8.2.dop}) is characterized by significantly larger value of $\alpha^{'}_{\mathbf{P}}(0)$ than the general version of the Regge-eikonal model. For the latter case $\alpha^{'}_{\mathbf{P}}(0)$ agrees well, especially at $\sqrt{s_{\min}}=30$ GeV, with the  "nominal" value. As seen from Tables \ref{tab:8-1add} and \ref{tab:8-1}, the simultaneous fit with one-Pomeron approach for the slope provides the same values of $\alpha^{'}_{\mathbf{P}}(0)$ as well as the single fit of the slope energy dependence by (\ref{eq:8.2.dop}). $B(s)$ has been calculated with Eq. (\ref{eq:8.2.dop}) and parameters were obtained from simultaneous fit with general Regge-eikonal relations (\ref{eq:8.3}), (\ref{eq:8.4}) and are shown in Table \ref{tab:8-1} at $\sqrt{s_{\min}}$ under consideration. In Fig. \ref{fig:4-2017}b results are shown by the dotted curve for $\sqrt{s_{\min}}=40$ GeV. There is a noticeable discrepancy between the dotted curve and both the experimental data and the results of simultaneous fit for $\mathcal{G}_{pp}$ with general Regge-eikonal formula (\ref{eq:8.4}) starting with RHIC energies $\sqrt{s} \gtrsim 100$ GeV. Furthermore, this difference increases with growth of collision energy $\sqrt{s}$. Therefore Fig. \ref{fig:4-2017}b clearly indicates the importance of multi-Reggeon effects at high energies. It is amusing, however, that these multi-Pomeron contributions mimic, at available energies, a power like form.

Below the simultaneous fit within Regge-eikonal approach with general Eq. (\ref{eq:8.4}) for the slope is considered. The Regge-eikonal model curves agree with experimental points
quite reasonably (Fig. \ref{fig:4-2017}). Fit parameters change
weakly at a noticeable improvement of the fit quality with growth of
$s_{\min}$. On the other hand, the approximations (\ref{eq:8.5})
describe experimental data for the set $\mathcal{G}_{pp}$ at
$\alpha^{'}_{\mathbf{P}}(0)$ which is significantly smaller than
that for Regge-eikonal model for corresponding energies while fit
qualities are almost the same for the two models under consideration
for narrower range $\sqrt{s} \geq 40$ GeV. Thus the fit quality
alone obtained for the energy domain $\sqrt{s} \geq 40$ GeV does
not allow us to give preference to either of the two models. The
above conclusion is illustrated by the Fig. \ref{fig:4-2017}. As
seen, the curves obtained from simultaneous fits within the
Regge-eikonal model (solid lines) and with help of approximations
(\ref{eq:8.5}) presented by dashed lines show very close behavior
up to the highest available energy $\sqrt{s} \approx 100$ TeV for
$\sigma_{\footnotesize\mbox{tot}}$ (Fig. \ref{fig:4-2017}a) and
for the slope (Fig. \ref{fig:4-2017}b) in the fitted energy domain
$\sqrt{s} \geq 40$ GeV.
As seen from Table \ref{tab:8-2}, the simultaneous fit of the
set $\mathcal{G}_{pp}$ by Eqs. (\ref{eq:8.5}) provides the
$c_{2}/8\pi b_{2}=1.16 \pm 0.03$ $(1.26 \pm 0.03)$ at
$\sqrt{s_{\min}}=30$ $(40)$ GeV. Thus the ratio $c_{2}/8\pi b_{2}$
exceeds the asymptotic level ($=1$) at more than 5 (8) standard
deviations (s.d.) at $\sqrt{s_{\min}}=30$ $(40)$ GeV. The only
advantage of Eqs. (\ref{eq:8.3}) and (\ref{eq:8.4}) is that expressions for
$\sigma_{\footnotesize\mbox{tot}}(s)$ and $B(s)$ are derived
consistently from the basics of the Regge-eikonal framework and
allow to falsify ("in the Popper sense") its premises while Eqs.
(\ref{eq:8.5}) seem to be a free phenomenological invention (although with a carefully chosen correct ratio between the coefficients before $\log^{2}s$ in expressions for $B(s)$ and $\sigma_{\footnotesize\mbox{tot}}$) the
failure of which does not entail any disastrous consequences.

\begin{figure}[!h]
\begin{center}
\includegraphics[width=12.0cm,height=10.8cm]{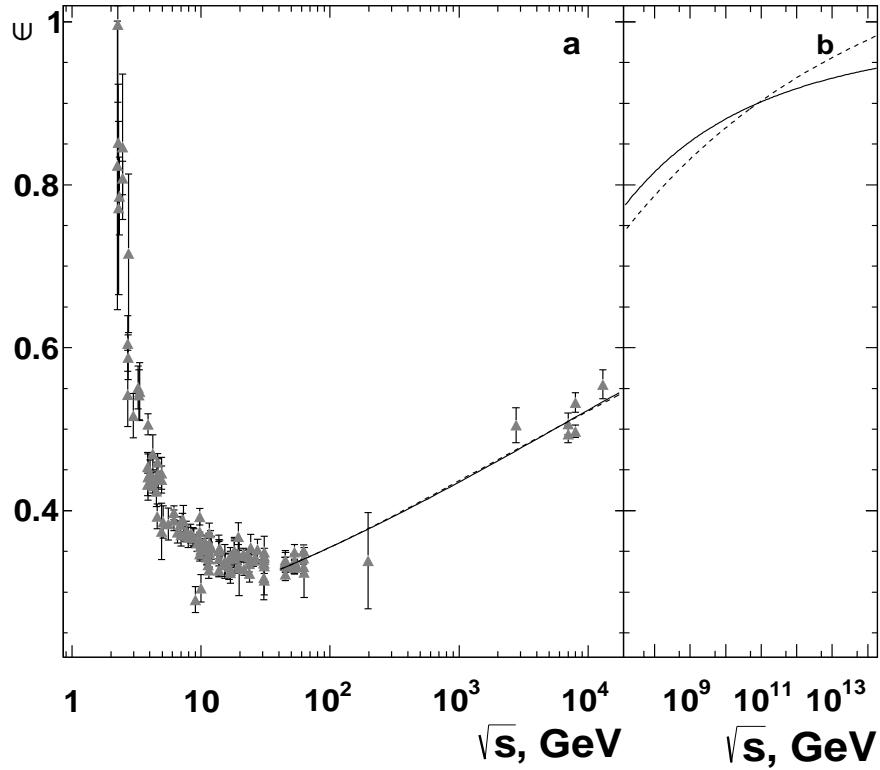} 
\end{center}
\caption{\textit{Energy dependence of $\varepsilon$  in the experimentally available energy domain (a) and for very high energies (b). The solid curve is obtained within the Regge-eikonal model and dashed curve is from Eqs. (\ref{eq:8.5}) with fit parameters for range $\sqrt{s} \geq 40$ GeV.}} \label{fig:6-2017}
\end{figure}

As was said above, the "asymptopia" would mean the energy region where
\begin{center}
$B(s) \gg r_{0}^{2}$.
\end{center}
It is true that the sign "$\gg$" looks rather vaguely but, nevertheless, from our formula we can see that even a modest condition
\begin{center}
$ B(s) = 3 r_{0}^{2}  $
\end{center}
needs the huge energy because even at $\sqrt{s}= 10$ PeV the ratio
$B(s) / (3 r_{0}^{2})$ is only $ 0.85 $.

Another indication of our
remoteness from the "asymptopia" is the value of the ratio
$\epsilon(s) \equiv \sigma_{\footnotesize\mbox{tot}}(s)/8\pi B(s)$
which should be asymptotically near to 1. In Fig. \ref{fig:6-2017} the
energy dependence of $\epsilon$ is shown for
experimentally available energy range (a) as well as in the domain of
very high energies (b). In the first case the estimations
calculated with help of measured
$\sigma_{\footnotesize\mbox{tot}}$ and $B$ at corresponding
energies are shown by symbols and smooth curves are deduced within
the Regge-eikonal model (\ref{eq:8.3}), (\ref{eq:8.4}) and Eqs.
\ref{eq:8.5} with parameter values obtained from simultaneous fits
at $\sqrt{s_{\min}}=40$ GeV and shown in Tables \ref{tab:8-1},
\ref{tab:8-2} respectively. At energies $\sqrt{s} > 10$ PeV only
phenomenological curves are shown in Fig. \ref{fig:6-2017}b and
exact relation (\ref{eq:8.3.dop2}) with parameter values from
Table \ref{tab:8-1} is used instead of approximate function
(\ref{eq:8.4}). As expected, the Regge-eikonal model and
approximations (\ref{eq:8.5}) agree quite reasonably with
experimental estimations for $\epsilon(s)$ at $\sqrt{s} \geq 40$
GeV and show very close behaviour up to the $\sqrt{s} \sim 100$
TeV. One can note that the experimental points and
phenomenological curves are far from the asymptotic level
$\epsilon = 1.0$ at experimentally available energies. In
a very high energy domain the both models provide the continuous
increase of $\epsilon(s)$. But the approximations (\ref{eq:8.5})
lead to some faster growth and, as a consequence, there is a
noticeable excess with respect to the Regge-eikonal
model at GUT energies $\sqrt{s}
> 10^{12}$ GeV (Fig. \ref{fig:6-2017}b). The study of the set of
the scattering parameters $\mathcal{G}_{pp}$ as well as of one of
the asymptotic signatures $\epsilon$ exhibits a very close
agreement between Eqs. (\ref{eq:8.3}), (\ref{eq:8.4}) and
approximation functions which are $\propto \ln^{2}(s/s_{0})$ at $s
\to \infty$ at least up to the energies $\mathcal{O}(100)$ TeV.
The noticeable difference between two approaches -- Regge-eikonal
model and approximations (\ref{eq:8.5}) -- can be expected only at incredibly
high energies $\sqrt{s} \geq 10$ PeV.

One can emphasize that in this paper we used a simple model (fairly sufficient for our limited goals) in which Regge poles are only exchange objects. Attempts to account  the $t$-channel unitarity led to a more sophisticated theoretical scheme, so-called  “Reggeon field theory” (RFT) in which reggeons are considered as quasi particles in an analog effective 2+1 field theory containing reggeon interaction terms (though in quite arbitrary Lagrangians). Interactions of reggeons  generally change the free reggeon exchange results.  Nonetheless, in \cite{PRD-85-094007-2012} authors on the basis of some simplified version of the RFT came to the conclusion that the “black disc regime” which is often associated with the onset of “asymptopia” should begin at energies higher than 57 TeV. This conclusion does not contradict our estimates. Other details of this scheme concern mostly the diffractive dissociation processes which were not the subject of the present work.

\section*{Conclusions} \label{sec:0}
 In this paper we had no task to achieve the best description of the data, so we limited ourselves with simply treated models designed for clear exhibiting our main observations and formulate physical corollaries. E.g., we completely ignored the real part of the scattering amplitude which as shown, for instance, in \cite{Ptr}, can sometimes play a crucial role. We hope, nonetheless, that such omissions cannot spoil the qualitative value of our main results.

Here they are.

Up to the LHC energies the proper sizes of protons cannot be neglected and they make a significant contribution in the size of the interaction region.

The very notion of the effective size depends on the process.

The juxtaposition of the interaction radius observed at the highest achieved by present energies (LHC) with the proper nucleon sizes shows that we are still extremely far from the "asymptopia', the region where some known asymptotic relations should hold.

The "technical" reason is that the effective evolution parameter characteristic for the Regge-eikonal approach is extremely slow function of the collision energy.

We do not want to create any pessimism in the reader regarding the unattainability of the "asymptopia" (quite possibly, a boring territory).

We would like to emphasize once more that it was not our aim, in this paper, to give an ideally accurate description of the characteristics in question, i.e. the slope and the total cross section, limiting ourselves with very minimal means to clarify and illustrate conceptual points.

Certainly, a genuine  paramount task is by no means an arbitrarily thorough description of a limited group os scattering characteristics but it rather would be a statistically well sounded and physically motivated description of the whole set of dynamically interrelated processes: elastic diffraction scattering, single- and double inelastic diffractive dissociation and more subtle subjects like central diffraction. The present paper shows very clear that the use of limited sets of observables (here total cross-section and slope) cannot discriminate even between models  quite different ideologically.

An ultimate goal (or, at least, one of the most important goals) is to use the best descriptions of the mentioned "complete set" of observables for extracting the properties of the fundamental entities in the realm of diffractive scattering, those of Reggeons \cite{Ryu}. These dreams are well supported by the newest observations in the diffraction region which are being obtained at the LHC and which give a rich food for those who are interested in real physics.

\section*{Acknowledgements} \label{sec:0}
The authors are grateful to A. M. Chebotarev, A. A. Godizov, V. V. Ezhela, A. V. Kisselev, A. K. Likhoded, R. A. Ryutin, A. P. Sa\-mo\-khin, S. M. Troshin and N. P. Tkachenko for useful discussions and stimulating criticism. The authors are indebted to the referee who has drawn our attention to the papers \cite{PRD-85-094007-2012,PRD-81-114028-2010}. V. A. P. is supported by the RFBR Grant 17--02--00120. The work of V. A. O. was supported partly by NRNU MEPhI Academic
Excellence Project (contract No 02.a03.21.0005, 27.08.2013).

\end{document}